\documentclass[aps,prl,twocolumn,amsmath,amssymb,nofootinbib,superscriptaddress,floatfix,reprint,longbibliography]{revtex4-2}
\usepackage{graphicx}
\usepackage{latexsym}
\usepackage{amsmath}
\usepackage{amsfonts}
\usepackage{amssymb}
\usepackage{bm}
\usepackage{color}
\usepackage{txfonts}
\usepackage{float}
\usepackage{url}
\usepackage[colorlinks=true, urlcolor=blue, linkcolor=blue, citecolor=blue]{hyperref}
\usepackage{ulem}
\usepackage{physics}
\DeclareUnicodeCharacter{03B4}{\ensuremath{\delta}}

\usepackage{tikz}
\newcommand*\circled[1]{\tikz[baseline=(char.base)]{
            \node[shape=circle,draw,inner sep=2pt, scale=0.8] (char) {#1};}}

\begin{document}
	\newcommand{\pprl}{Phys. Rev. Lett. \ }
	\newcommand{\pprb}{Phys. Rev. {B}}

\title{Coulomb-Driven Electronic Textures in Lightly Hole-Doped Cuprates}    

\author{Wei-Wei Yang}
\thanks{These authors contributed equally.}
\affiliation{Beijing National Laboratory for Condensed Matter Physics and Institute of Physics,
	Chinese Academy of Sciences, Beijing 100190, China}

\author{Yang Shen}
\thanks{These authors contributed equally.}
\affiliation{School of Physical Science and Technology, ShanghaiTech University, Shanghai 201210, China}

\author{Yuxin Wang}
\affiliation{Beijing National Laboratory for Condensed Matter Physics and Institute of Physics,
	Chinese Academy of Sciences, Beijing 100190, China}
\affiliation{School of Physical Sciences, University of Chinese Academy of Sciences, Beijing 100190, China}

\author{Pengfei Li}
\affiliation{Beijing National Laboratory for Condensed Matter Physics and Institute of Physics,
	Chinese Academy of Sciences, Beijing 100190, China}
\affiliation{School of Physical Sciences, University of Chinese Academy of Sciences, Beijing 100190, China}

\author{Mingpu Qin}
\email{qinmingpu@sjtu.edu.cn}
\affiliation{Key Laboratory of Artificial Structures and Quantum Control (Ministry of Education), School of Physics and Astronomy, Shanghai Jiao Tong University, Shanghai 200240, China}

\author{Kun Jiang}
\email{jiangkun@iphy.ac.cn}
\affiliation{Beijing National Laboratory for Condensed Matter Physics and Institute of Physics,
	Chinese Academy of Sciences, Beijing 100190, China}
\affiliation{School of Physical Sciences, University of Chinese Academy of Sciences, Beijing 100190, China}

\date{\today}

\begin{abstract}
Motivated by recent experiments on  Ca$_2$CuO$_2$Cl$_2$, we present a comprehensive study of the interplay between dopants and localized Zhang–Rice singlets in lightly hole-doped cuprates.
Owing to the poorly screened Coulomb potential, the interactions between dopants and localized Zhang–Rice singlets favor asymmetric electronic configurations characterized by effective electric `dipoles'.
Using the MPS-based Lanczos method, we show that these localized holes give rise to in-gap bound states within the charge-transfer gap. When tunneling matrix-element effects are included, the bound states exhibit a characteristic $4a_0 \times 2a_0$ internal structure, consistent with experimental observations. 
Our results suggest that Coulomb interactions between dopants and localized Zhang–Rice singlets, together with tunneling matrix-element effects, provide a microscopic mechanism for the observed \(4a_0\) electronic structures.

\end{abstract}
\maketitle

{\em Introduction.}---
Carrier doping has the profound ability to fundamentally alter the physical properties of an insulator, transforming it from a material with negligible conductivity to one that exhibits a wide range of new functionalities.
For example, doping the semiconductor is one essential step in the semiconductor industry to achieve PN junctions and other functional devices \cite{shockley1950electrons,neamensemiconductor}.
Similarly, in strongly correlated materials, doping is essential for realizing emergent phases of matter. A notable example is the doping of Mott insulators (or charge transfer insulators)\cite{Zannen_PhysRevLett.55.418,khomskii2014transition,Kowalski_PNAS_2021}, which is widely regarded as the initial step toward achieving high-temperature superconductivity \cite{doping_mott,Anderson_2004,keimer2015quantummatter}. Although this doping has been discussed for many decades, the doping process remains poorly understood due to the complexity of electron correlation and limited characterization \cite{imada_RevModPhys.70.1039,Dagotto_RevModPhys.66.763,Shen_RevModPhys.75.473,Phillips_RevModPhys.82.1719}.

Recently, a scanning tunneling microscopy (STM) study on the lightly doped cuprate Ca$_2$CuO$_2$Cl$_2$ (CCOC) provides new insights into this old fundamental problem \cite{ye2023visualizing,yayu_wang_2025}, where a localized Zhang-Rice (ZR) singlet is identified \cite{li2024boundstatesdopedcharge,xia2024unveilingstripeshapedchargemodulations}. This localized ZR shows similar behavior to the impurity states with hydrogen-like wavefunction in doped semiconductors \cite{shockley1950electrons,KOHN1957257}. 
The major difference between these two cases is the Coulomb screening.
For the Mott insulator, the Coulomb screening is quite weak. Hence, the bound energy is much stronger and the bound radius becomes much shorter. The poor screened Coulomb potential is also the reason for electronic inhomogeneity in cuprates \cite{pan_inhomogeneity,WangZQ_PhysRevB.65.064509}. 
As a consequence, how the localized carriers behave under poor screening of the Coulomb potential remains an essential question.

\begin{figure}[t]
	\begin{center}
		\includegraphics[width=2.8in]{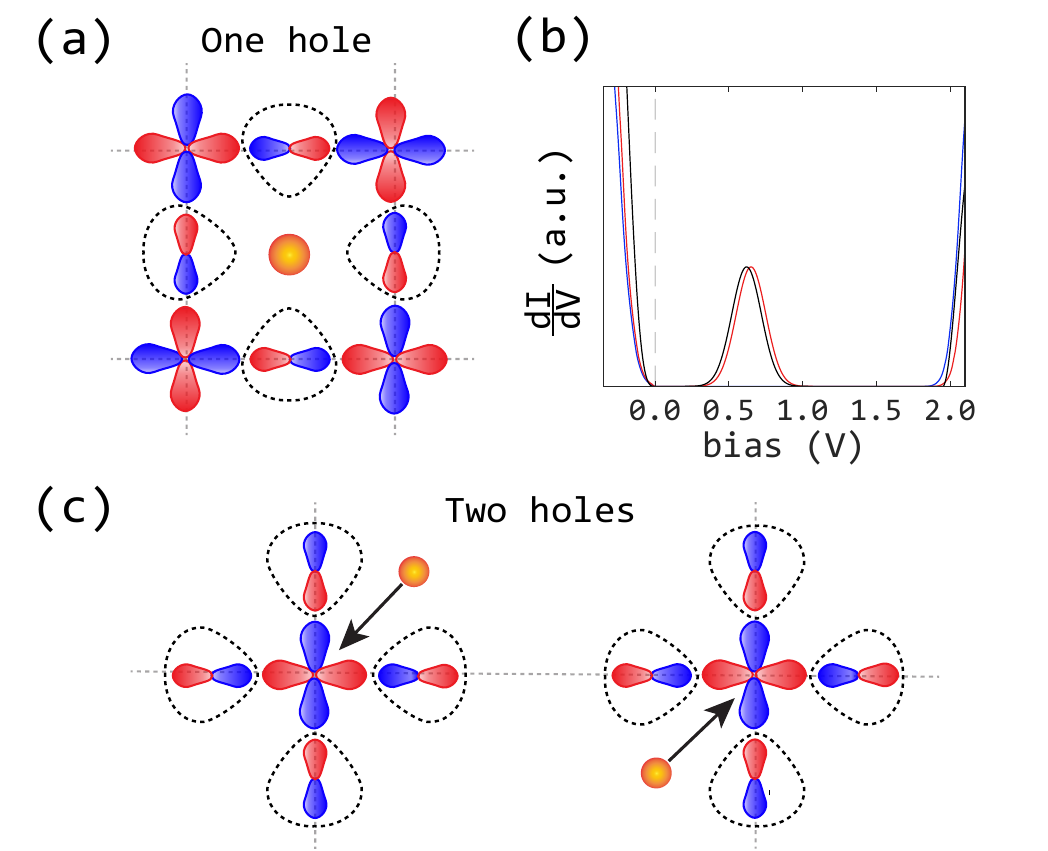}
		\caption{
			 Schematic representations and tunneling spectroscopy of doped CCOC. (a) Schematic diagram for the 1-hole doping scenario on the pristine Mott insulator, where the hole is uniformly distributed within the O p-orbitals around the doped Na ions in CCOC. (b)  The schematic scanning tunneling spectroscopy density of states in lightly Na doped cuprates Ca$_2$CuO$_2$Cl$_2$. The blue curve shows the spectrum far from the Na dopant. The black and red curves correspond to the Na-centered symmetric state (a) and the Cu-centered four-lobe clover state (c), respectively. Both exhibit a bound state within the charge-transfer gap at $E_b\sim0.6$--$0.7\,\mathrm{eV}$\cite{ye2023visualizing}.  (c) Schematic diagram 2-hole doping scenario on the pristine Mott insulator, where the holes are distributed within the O p-orbitals around selected Cu sites near the doped Na ions in CCOC. In diagrams (a) and (c), yellow represents the doped Na in CCOC, while blue and red indicate the relative phases of the wave functions for O's 2p orbitals and Cu's 3d orbitals, respectively.
			\label{fig1}}
	\end{center}
\end{figure}

The hint for this question also comes from the CCOC experiments \cite{ye2023visualizing}.
At the extremely dilute dopant case, 
the scanning tunneling microscope (STM) results show a spatial symmetric distribution, as illustrated in Fig.~\ref{fig1}(a). 
The dominant tunneling hotspot surrounds the central dopant Na site.  As the concentration of dopants in the system increases, another unique `four-lobe' clover pattern emerges around the central Cu site instead, as depicted in Fig.~\ref{fig1}(c). 
In both cases, a bound state around $E_b\sim0.6$--$0.7\,\mathrm{eV}$ shows up at the charge transfer gap (see Fig.~\ref{fig1}(b)), which thus has been attributed to a localized ZR singlet \cite{li2024boundstatesdopedcharge}.
Interestingly, a $4a_0$ periodic pattern is also identified in the multiple dopants, which is also proposed to be the fundamental unit for the cuprate Cooper pairing \cite{ye2023visualizing,yayu_wang_2025}.
In this work, we will show that the localized ZR and the Na dopant form an effective electric ``dipole", leading to the `four-lobe' pattern.
We also find that dopant–hole interactions favor specific electronic configurations. 
The interaction between these electric ``dipoles" leads to the inner structure of hole dopants. We propose that this inner structure may be the reason for 4a$_0$ pattern observed in STM.
We want to emphasize that although we use CCOC as an example, our conclusion is general for cuprates and other doped-Mott insulators.





{\em Ground state of the two-hole situation.}--- Ca atoms are substituted by Na atoms in CCOC \cite{ye2023visualizing,ccoc_0,ccoc_1,ccoc_2}.
Since Na has one fewer valence electron than Ca, it acts as an electron acceptor, introducing one hole into the system. In the extremely dilute case, the dopants are well separated. 
The doped hole is attracted to the Na dopant in a manner similar to how an electron is bound to a hydrogen atom, as depicted in Fig.\ref{fig1}(a). This feature is similar to the boron doping silicon \cite{shockley1950electrons,neamensemiconductor}.
However, as the dopant density increases, the charge of each dopant begins to sense the presence of additional Coulomb potentials as other dopants come closer. This additional potential disrupts the symmetric distribution of charges, leading to an asymmetric configuration, as shown in Fig.\ref{fig1}(c). 
The asymmetric configuration plays the same role as an effective electric dipole, as illustrated by the arrow pointing from the negative charge center to the positive hole center, i.e., from Na to Cu in  Fig.\ref{fig1}(c).
Clearly, this dipole configuration saves more Coulomb energy than Fig.\ref{fig1}(a).

Mathematically, the above statement can be obtained by evaluating the total energy by
\begin{equation}
\begin{aligned}
E_{total}=V_{ii}+V_{ie}+V_{ee}+E_{kin}+E_{other} ,
\end{aligned}
    \label{eq:energy}
\end{equation}
where $V_{ii}$ is the ion-ion interaction between two Na$^{1+}$ ions, $V_{ie}$ is the interaction between Na$^{1+}$ ions and the doped holes, and $V_{ee}$ is the electron interaction between two holes. We simply ignore the kinetic energy $E_{kin}$ owing to the localization. $E_{other}$ is the other energy of charge transfer insulator background.
To simplify our calculation, we adopt the Hatree-Fock approach for small molecules in quantum chemistry \cite{szabo1996modern,Slater_PhysRev.81.385,KOHN1957257}. 
One central assumption for this approach is that the ground state wavefunction is formed from localized ZR singlets and AFM background \cite{li2024boundstatesdopedcharge}.


The $V_{ii}$ has been implemented by the standard DFT packages. We use the Quantum Espresso package, where the $V_{ii}$ is calculated by the Ewald contribution \cite{QE_package}. The $V_{ie}$ is calculated by
\begin{equation}
\begin{aligned}
V_{ie} = - e^2 \int dr_1 dr_2 \rho_{Na}(r_1) \psi_{i}^{*} (r_2)\psi_{i}(r_2),
\end{aligned}
    \label{eq:interaction1}
\end{equation}
where the Na$^{1+}$ is taken as a point charge ($\rho_{Na}(r)=\delta(r-r_{Na})$). $\psi_{i}(r)$ is the wavefunction of ZR hole at oxygens, which is a function of the real-space coordinate $r$, and the index $i$ labels the hole hosted by the $i$-th Na$^+$ ion.
The $V_{ee}$ interaction is represented by the Hatree term $V_{H}$ and the Fock term $V_{F}$ as
\begin{equation}
\begin{aligned}
V_{H} &= + e^2 \int dr_1 dr_2 \psi_{i}^{*} (r_1)\psi_{i}(r_1) \frac{1}{|r_1-r_2|}\psi_{j}^{*} (r_2)\psi_{j}(r_2) \\
V_{F} &= - \frac{e^2}{2} \int dr_1 dr_2 \psi_{i}^{*} (r_1)\psi_{j}(r_1) \frac{1}{|r_1-r_2|} \psi_{j}^{*} (r_2)\psi_{i}(r_2). \\
\end{aligned}
    \label{eq:interaction2}
\end{equation}
By applying the above formulas to our case, we derive the configurations of holes and Na dopants that possess the lowest energy in both single-hole and double-hole scenarios.

\begin{figure}[t]
	\begin{center}
		\includegraphics[width=2.7in]{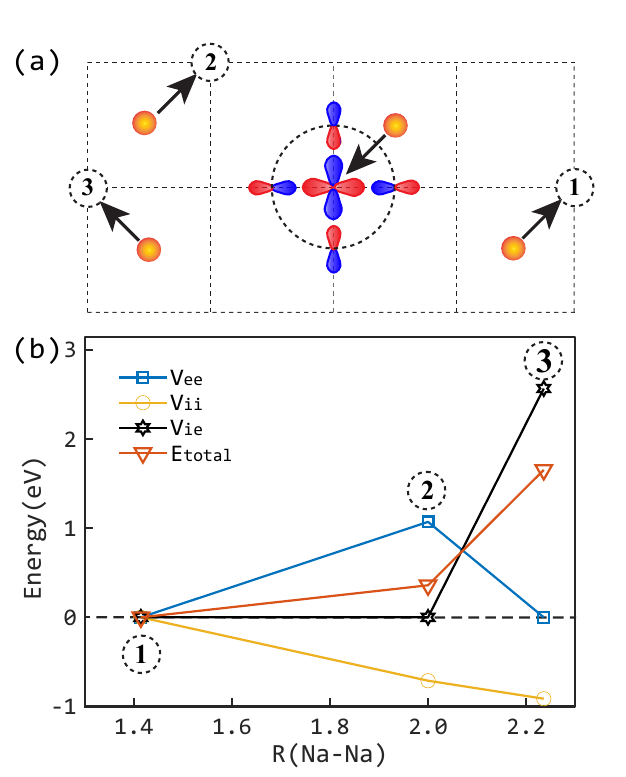}
		\caption{ Potential Configurations and Corresponding Energies for a Two-Hole Doping Scenario.(a) We explore various configurations of two doped Na ions at increasing distances from one another, specifically at distances of \( r = \sqrt{2}a \), \( 2a \), and \( \sqrt{5}a \), labeled as configurations 1, 2, and 3, respectively. In the figure, we present only the lowest energy configurations. The dashed circles indicate the positions of the second hole in each scenario.(b) Corresponding to the configurations depicted in (a), we present the associated energies. The total energy (represented by purple triangles) comprises contributions from electron-electron interactions (blue squares), electron-ion interactions (black hexagons), and ion-ion interactions (yellow circles).
			\label{fig2}}
	\end{center}
\end{figure}

For the single-hole scenario, the doped hole is simply as in Fig.\ref{fig1}(a) resulting in a zero electric dipole moment.
For the two-hole scenario, how to arrange the dopants becomes crucial.
We list three possible dipole configurations in Fig.\ref{fig2}(a). The distance between the two Na atoms is $\sqrt{2}a_0$, $2a_0$, and $\sqrt{5}a_0$, respectively. Then, we should obtain energy contours that take the same spirit as the Lennard-Jones potential.
However, it is important to emphasize that we exclude overlapping configurations of the two dipoles that are dominated by strong Coulomb repulsions, which lie beyond the scope of the current simplified approach.



The energy calculation results are presented in Fig. \ref{fig2}(b), with all energies referenced to configuration-\circled{1}. 
First, take a look at $V_{ii}$. As expected, the repulsive $V_{ii}$ decreases as the dopant distance $r$ increases, as illustrated by the yellow line. Then, the attractive $V_{ie}$ depends on both $r$ and the orientation of the dipole arrows, with configurations \circled{1} and \circled{2} exhibiting the same energy, while \circled{3} requires higher energy. Finally, configurations \circled{1} and \circled{3} share the same repulsive $V_{ee}$, whereas \circled{2} incurs greater repulsion energy. 
The Hartree and Fock terms in $V_{ee}$, 
$V_H$ and $V_F$, are determined by the positions of the electronic wavefunctions of the two dipoles. $V_H$ contributes positively to the total energy, while $V_F$ leads to a reduction in energy. For configurations \circled{1} and \circled{3}, their $V_{ee}$ are identical, with $V_H=1.43eV$ and $V_F=-0.18eV$. However, in the case of \circled{2}, the electrons are closer together, resulting in larger absolute values for both the Hartree and Fock energies. The increase in the Hartree term is significantly greater than that in the Fock term, leading to a higher electron-electron interaction energy.
Summing all these terms, the total energy for each configuration is plotted as the red line. Among the configurations, \circled{1} has the lowest total energy, as it is favored in both the $V_{ie}$ and $V_{ee}$ interactions, and this advantage holds even when the repulsive $V_{ii}$ is considered.
This effect arises because the dipoles possess orientational degrees of freedom, and we note that $V_{ie}$ dominates the total energy, followed by $V_{ee}$, and finally $V_{ii}$. In the ground state \circled{1}, the antiparallel alignment of the dipoles maximizes the attractive interactions between Na$^+$ and electron, while simultaneously maximizing the distance between the two electronic wavefunctions (as illustrated in Fig. \ref{fig2}(a)). Consequently, in doped CCOC, Na substituents at specific positions and separations exhibit the lowest energy in this particular configuration \circled{1}.

The single-hole scenario (Fig.~\ref{fig1}(a)) is also carefully studied, where dipoles are assumed to be infinitely separated.
The stability of the double-hole system becomes evident when compared to the single-hole case.
This energy advantage arises because the double-hole system benefits from a strong, additional attractive interaction $V_{ie}$ (with $\delta V_{ie} = -11.25$ eV) between a Na$^+$ ion and a hole on a neighboring dipole. Although this configuration also introduces repulsive interactions ($\delta V_{ee} = 2.37$ eV and $\delta V_{ii} = 1.21$ eV), the energy gain from $V_{ie}$ far exceeds the energy cost from the repulsions. Furthermore, the dipole structure itself plays a key role in significantly mitigating the repulsive interaction $V_{ee}$ between the doped holes. The net energy reduction conclusively explains the stability of the asymmetric configuration observed in STM experiments \cite{ye2023visualizing}.

\begin{figure}
	\begin{center}
		\includegraphics[width=2.6in]{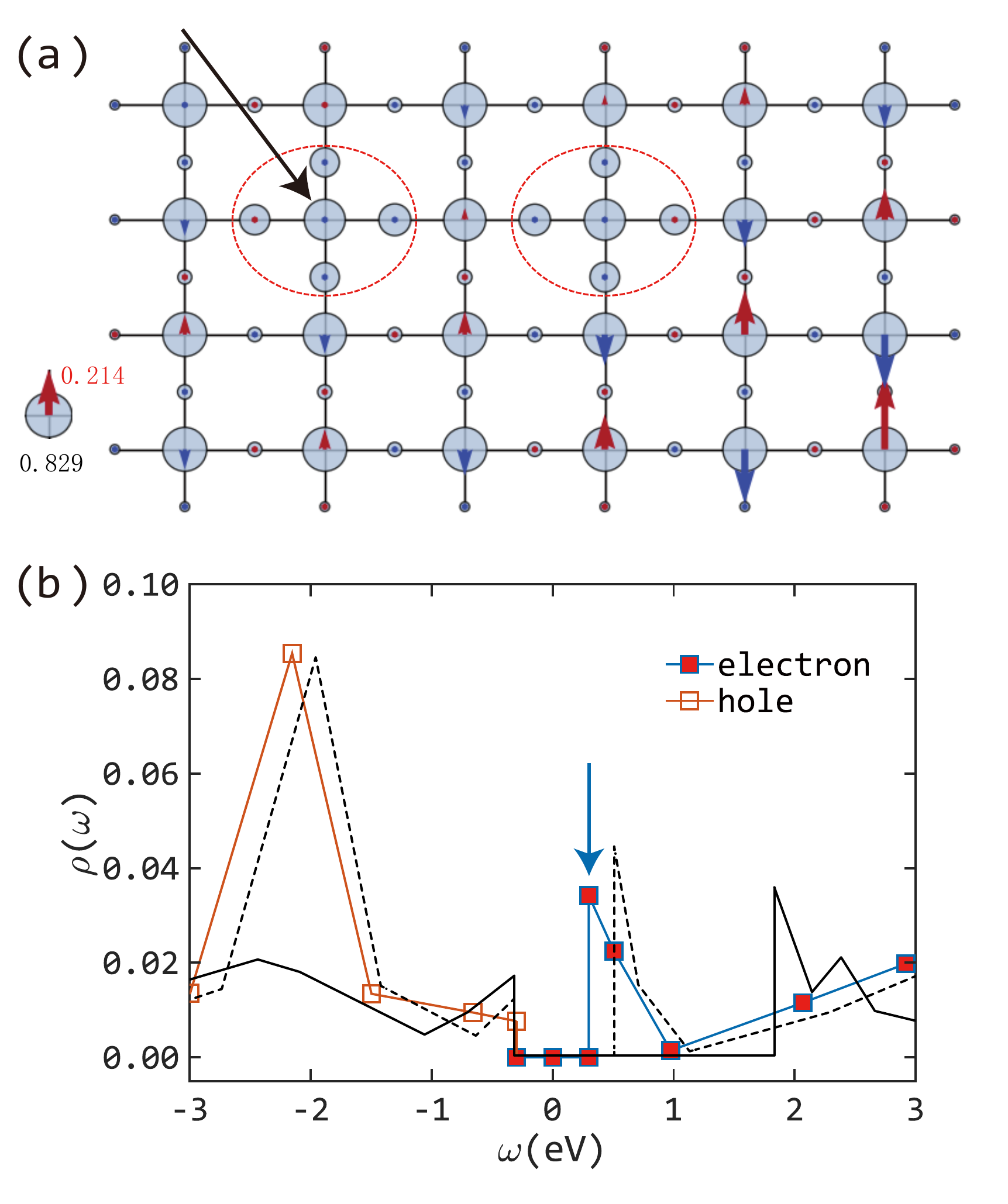}
		\caption{
Ground-state distributions and local spectra for a system comprising $6\times4$ CuO$_2$ unit cells, calculated using the MPS-based Lanczos method. (a) The spin and hole distribution for the ground state with 2 extra hole doping. The doped holes mainly reside on the eight oxygen sites surrounding the two Cu centers marked by the dashed ovals, where the local oxygen onsite energies are lowered. (b) The local density of states (LDOS), $\rho(\omega)$, at the Cu site indicated by the black arrow in (a) is plotted for the two-hole-doped ground state as a function of energy, $\omega$. The filled (open) markers represent result for processes adding one electron (hole) to the two-hole doping ground state. For reference, we also plot the Cu-site LDOS for the half-filled charge-transfer insulator (solid black line) and the one-hole doped case (dashed black line).
			\label{fig3M}}
	\end{center}
\end{figure}   

{\em tunneling spectrum of STM.}---
After identifying the possible configurations, we proceed to a detailed analysis of their tunneling spectra. Since our interest lies in the low-energy electronic excitations, we restrict our consideration to the electronic sector and neglect the ionic contribution. In this limit, the Hamiltonian reduces to the three-band Emery model with two localized Zhang–Rice holes, that is, we eliminate electronic counterpart residing on Na+ but assume the existence of a local dipole in the ground state. \cite{Emery_PhysRevLett.58.2794,Zhang_Rice,Feiner_PhysRevB.45.7959}.
In addition to the standard Emery terms, the localization potential arising from dopants can be approximated as
\begin{equation}
    H_{\mathrm{loc}} = E_{\mathrm{loc}}
    \sum_{i_0 \in \mathcal{I}_{\mathrm{ZR}}}
    \sum_{\delta,\sigma}
    p_{i_0\delta\sigma}^\dagger p_{i_0\delta\sigma}
    \label{eq:Hloc}
\end{equation}
where $\mathcal{I}_{\mathrm{ZR}}=\{i_1,i_2\}$ contains the two Cu sites hosting the localized Zhang--Rice singlets, marked by dashed ovals in Fig.~\ref{fig3M}(a).
For each center $i_0$, $\delta$ labels its four nearest-neighbor oxygen sites, and $\sigma\in\{\uparrow,\downarrow\}$ labels the spin.
The operator $p_{i_0\delta\sigma}^{\dagger}$ creates a hole in the corresponding O $2p_x$ or $2p_y$ orbital.
The dopant-induced local potential $E_{\mathrm{loc}}$ shifts the oxygen onsite energies equally for both spin states.
For the configuration in Fig.~\ref{fig3M}(a), the two oxygen neighborhoods are disjoint, so the sum covers eight distinct oxygen sites.
The three-band parameters used are adapted from an ab initio calculation for CCOC with $t_{pd}=0.78eV$, $t_{pp}=0.15eV$, $\tilde{\epsilon}_{p}-\tilde{\epsilon}_{d}=3.25eV$, $U=9.5eV$, $E_{loc}=-2t_{pd}$ \cite{li2024boundstatesdopedcharge,kent_PhysRevB.78.035132}.

Since, in both single-hole and double-hole scenarios, the energy scale of the CT gap ($\Delta_{CT} \sim 2 eV$) and the bound state energy ($\Delta \sim 0.6-0.7 eV$) are significantly larger than the exchange energy ($J \sim 0.14 eV$) and the hopping scale ($t \sim 0.35 eV$) \cite{kim_PhysRevLett.80.4245,Chernyshev_PhysRevB.50.13768} and we are studying local physics, the finite size effect should be small \cite{li2024boundstatesdopedcharge}.

\begin{figure}
	\begin{center}
		\includegraphics[width=2.6in]{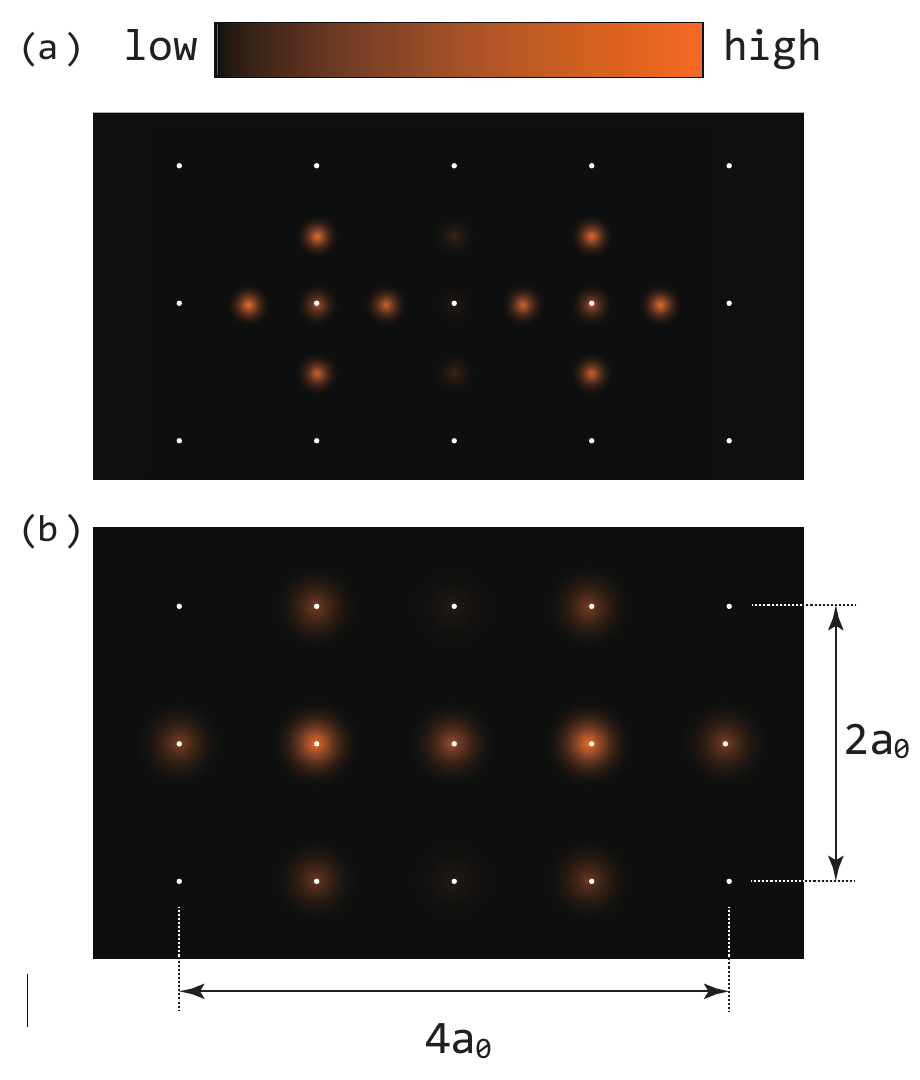}
		\caption{
STM result $\rho(\omega)$ for a system consisted of $6\times4$ CuO$_2$ unit cells from MPS-based Lanczos method.
Local density of states for the in-gap bound state identified in Fig3.~(b) with blue arrow are shown both without (a) and with (b) the inclusion of the tunneling matrix effect. The white dot pattern signifies the positions of Cu sites in CCOC, analogous to Cl atom locations. In (a), the DOS map assumes direct coupling of the STM probe to Cu or O atoms. Notably, in (b), the STM map evaluated above the Cu sites displays a stripe pattern with an approximate width of 4$a_0$.
			\label{fig4M}}
	\end{center}
\end{figure}   

We employ an MPS-based Lanczos approach to study a system comprising ($6 \times 4$) CuO$_2$ unit cells with open boundary conditions, as illustrated in Fig.~\ref{fig3M}(a). Technical details of the MPS-based Lanczos method can be found in Ref.~\cite{PhysRevB.85.205119}, where post-reorthogonalization techniques are introduced to enhance numerical accuracy. Two additional holes are introduced into the system, with their pinning positions with $E_{loc}$ indicated by the dashed ovals in Fig.~\ref{fig3M}(a). To visualize the spin configuration, we further apply a magnetic pinning field ($h_m = 0.1t_{pd}$) at the Cu site in the lower-right corner.



From the doped-hole ground state, we extract the spin and hole density distributions at each site, as shown in Fig.~\ref{fig3M}(a). Within the dashed ovals, the oxygen sites exhibit significantly enhanced hole density compared to other oxygen sites, demonstrating that the doped holes are indeed localized in these regions, forming two Zhang–Rice singlets. The excitation spectrum is subsequently obtained via the Lanczos procedure. The local density of states at the Cu site indicated by the black arrow in Fig.~\ref{fig3M}(a) is presented in Fig.~\ref{fig3M}(b), revealing a finite energy gap of $0.60eV$ separating electron and hole excitations. For comparison, the local Cu spectrum of the charge-transfer insulator is shown by the solid black curve, exhibiting a much larger gap of $2.15 eV$. This establishes the presence of bound states within the charge-transfer gap. A similar feature has been reported in the one-hole case, where the gap is slightly larger $0.83eV$ \cite{li2024boundstatesdopedcharge}. Altogether, these results are consistent with experimental observations and with the localized Zhang–Rice singlets shown in Fig.~\ref{fig1}.




We next examine the spatial distribution of the DOS at the bound-state energy, as shown in Fig.~\ref{fig4M}. In Fig.~\ref{fig4M}(a), the DOS intensity is concentrated around the two pinning centers, particularly on the oxygen sites. This `four-lobe’ pattern provides further support for the localized Zhang–Rice singlet scenario.
In realistic STM measurements, however, it is essential to account for tunneling matrix-element effects \cite{li2024boundstatesdopedcharge,rice_tunnel,YanChen_PhysRevLett.97.237004}. The STM tip predominantly couples to the CCOC surface via the apical Cl $p_z$ orbital. Importantly, since the Cl sites are positioned directly above the Cu atoms, the overlap between $p_z$ and $d_{x^2-y^2}$ orbitals vanishes by symmetry. Incorporating this effect yields the STM map shown in Fig.~\ref{fig4M}(b), where the spatial distribution of the two-hole cluster becomes more extended, acquiring a characteristic $4a_0 \times 2a_0$ pattern.

Interestingly, the $4a_0 \times 2a_0$ pattern we obtain has also been observed experimentally in the stripe structures reported at 3\% doping in Ref.~\cite{ye2023visualizing}. The $4a_0$ modulation, in particular, has been widely documented in earlier STM studies \cite{4a0-Kapitulnik,4a0-Hanaguri,4a0-Yazdani,4a0-mesaros,yuyijun}. 
More recently, it has been proposed that a $4a_0 \times 4a_0$ plaquette hosting two holes constitutes a key element underlying cuprate superconductivity \cite{yayu_wang_2025,xia2024unveilingstripeshapedchargemodulations,Yao_PRL_2026_156001}. 
While our simplified calculation cannot be expected to reproduce all experimental details, our findings suggest that the essential origin of the $4a_0$ internal structure revealed in STM lies in the Coulomb-driven redistribution of localized holes. 
This electronic redistribution arises from 
the poorly screened Coulomb potential inherent to the doped Mott insulating state. Specifically, when dopant ions introduce holes into the insulating system, the dopant spatial arrangement is governed by Coulomb interactions. To minimize the total energy $E_{total}$, the doped holes remain influenced by the attractive potential of the dopants, leading to the characteristic internal structure observed in STM measurements.

{\em Conclusion.}---In summary, starting from the localized Zhang–Rice singlet picture, we systematically investigate the spatial arrangement of ions and holes in a charge-transfer insulator with two dopants introducing two additional holes.
Owing to the presence of other dopants, the dopants and holes form a `dipole' configuration to minimize the poorly screened Coulomb potential in the doped Mott insulating state. The resulting `dipole–dipole' interaction drives energetically favorable dipolar configuration
at the Na–Na distance of $\sqrt{2}a_0$, minimizing the total electronic energy. 
Using the MPS-based Lanczos method, we further identify localized ZR singlets that produce in-gap bound states with an energy gap of $0.60eV$, in contrast to the $2.1eV$ charge-transfer gap of the undoped system. When tunneling matrix-element effects are included, the spatial distribution of these bound states exhibits a characteristic $4a_0 \times 2a_0$ internal structure, consistent with the STM observations reported in Ref.~\cite{ye2023visualizing}. We propose that the origin of this $4a_0$ modulation lies in Coulomb-driven electronic configurations in the doped Mott insulating background.

\bibliography{main_v3}

\clearpage
\onecolumngrid
\begin{center}
\textbf{\large Supplemental Material: Coulomb-Driven Electronic Textures in Lightly Hole-Doped Cuprates}
\end{center}

\setcounter{equation}{0}
\setcounter{figure}{0}
\setcounter{table}{0}
\setcounter{page}{1}
\makeatletter
\renewcommand{\theequation}{S\arabic{equation}}
\renewcommand{\thefigure}{S\arabic{figure}}
\renewcommand{\thetable}{S\arabic{table}}

\onecolumngrid
\raggedbottom
\noindent
\begin{minipage}[t]{0.50\textwidth}
\vspace{0pt}
\fontsize{10}{11}\selectfont
{\em LDOS of the ground state calculated by ED}---
The basic electronic structure and other spectra can be understood qualitatively from a cluster's exact diagonalization.
In our exact diagonalization study, we consider the Cu-O cluster with three Cu atoms and their nearest O atoms, as shown in the inset of Fig.~\ref{figS1} (a).
The tunneling density of states (DOS) at $\mathbf{r}$ for STM following Fermi's golden rule can be written as
\begin{eqnarray}
    \frac{dI}{dV}(\mathbf{r},\omega) \propto \sum_{n} |\langle n| \phi_{\sigma}^{\dagger}(\mathbf{r}) |GS\rangle|^2 \delta(-\omega-E_n+E_{GS}) \nonumber \\
    +\sum_{m} |\langle m| \phi_{\sigma}(\mathbf{r})|GS\rangle|^2 \delta(-\omega-E_{GS}+E_{m})
    \label{dIdV}
\end{eqnarray}
where $\phi_{\sigma}^{\dagger}(\mathbf{r})$ will be equal to $ d_{i\sigma}^\dagger$ or  $p_{l\sigma}^\dagger$ depending on the position and $\omega$ is the energy in the unit of $eV$. $|n\rangle$ are adding hole excitation eigenstates with hole number $n>N$ and $|m\rangle$ are removing hole excitation eigenstates with hole number $m<N$. 
Note that we have switched the energy $\omega$ in the delta functions to the electron picture for experimental comparison. The resulting local spectrum is plotted in Fig. \ref{figS1} (a). The ground state of the above cluster contains $N=5$ total holes.
The gap $\Delta_{gap}$ between transitions $N \rightarrow N-1$ ($\omega>0$) and $N \rightarrow N+1$ ($\omega<0$) is approximately $0.48 eV$, attributed to the charge transfer process. This value is smaller than the $1.2 eV$ observed in the single-hole case and the $2.8 eV$ in the undoped scenario reported in previous studies \cite{li2024boundstatesdopedcharge}. This diminishment is owing to the overlapping between two localized holes at oxygens.

Then, we look at the spatial tunneling distribution at the first electron excitation state, labeled as bc in Fig.~\ref{figS1} (a).
The color bar of Fig.~\ref{figS1} (b-c) shows the intensity of tunneling DOS at each spatial site. 
The brightest spot in Fig.~\ref{figS1} (b) is the middle Cu indicating the first excited state containing the overlapping between two holes. 
On the other hand, we need to consider the tunneling matrix effect to compare with experiments. Thus, the STM spectral intensity can be obtained using the formula:
\begin{eqnarray}
    \phi_\sigma^\dagger(\mathbf{r}) =  \sum_\tau (-1)^{P_\tau} \braket{\mathbf{r}}{i,Cl} p^\dagger_{i,\tau,\sigma}
    \label{STM}
\end{eqnarray} 
where $\braket{\mathbf{r}}{i,Cl}$ is the overlap between STM tip and Cl $p_z$ state at site i. The term describes the tunneling to O $p_{x(y)}$ around the Cu at site i in which $(-1)^{P_\tau}=-1$ for $\tau=-x,-y$ and $(-1)^{P_\tau}=1$ for $\tau=x,y$ according to symmetry.
The tunneling DOS at the Cl (on top of Cu) sites are shown in Fig.\ref{figS1} (c).
The STM signal is strongest above the Cu centers hosting the localized Zhang--Rice singlets at this energy. Interestingly, we find the spatial distribution has an internal $4a$ structure along the direction connecting the two Cu centers while a $2a$ along the perpendicular direction.
This $4a$ feature is owing to the Cl $p_z$ tunneling into the O $p_x$ orbital.
The STM signals evaluated above the left and right edge Cu sites arise solely from the $p_x$ orbitals of the O atoms on their respective right and left sides. 
The STM signals evaluated above the upper and lower rows of Cu sites arise from the $p_y$ orbitals of the O atoms below and above, respectively, leading to slightly lower DOS intensities at these positions. These two observations are consistent with the experimental findings in Ref.~\cite{ye2023visualizing}.
\end{minipage}\hfill
\begin{minipage}[t]{0.47\textwidth}
\vspace{0pt}
\makeatletter\def\@captype{figure}\makeatother
\centering
\includegraphics[width=2.7in,trim=0 20bp 0 20bp,clip]{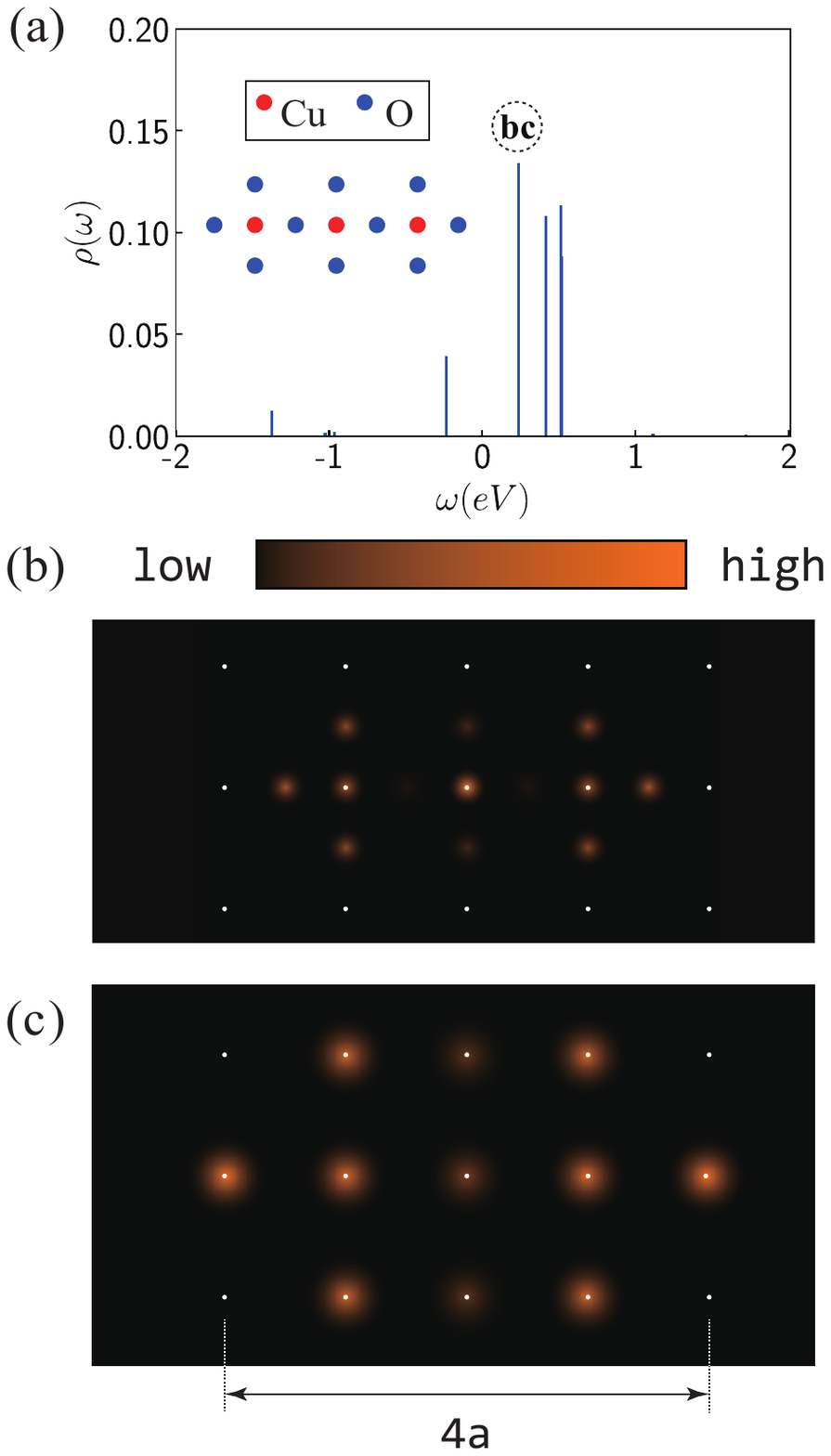}
		\caption{
(a) The tunneling density of states (DOS), $\rho(\omega)$, calculated for a two-hole doping scenario in the ground state, $|GS\rangle$, is plotted as a function of energy, $\omega$. The inset provides a schematic of the 13-site Cu-O plaquette, with red circles indicating Cu atoms and blue circles representing O atoms.
(b-c) Local density of states for the in-gap bound state identified in (a) are shown both without (b) and with (c) the inclusion of the tunneling matrix effect. The white dot pattern signifies the positions of Cu sites in CCOC, analogous to Cl atom locations. In (b), the DOS map assumes direct coupling of the STM probe to Cu or O atoms. Notably, in (c), the STM map evaluated above the Cu sites displays a stripe pattern with an approximate width of 4$a_0$.
			\label{figS1}}
	
\par\vspace{10pt}
\begingroup\fontsize{10}{11}\selectfont\rightskip=0pt\leftskip=0pt\parfillskip=0pt plus 1fil\relax
{\em Configuration with Two Adjacent Na Dopants.}---
In this section, we present another scenario not considered in the main text: the case in which two Na atoms substitute two nearest-neighbor Ca atoms. A schematic illustration of this two-adjacent-Na-dopant configuration is shown in Fig.~\ref{figS2}. This configuration was excluded from the main discussion because it leads to an unstable structure featuring two holes localized around one Na site, despite exhibiting a relatively low total energy. The corresponding tunneling density of states is presented in Fig.~\ref{figS3}.

The ground state of the above 10-site cluster contains $N=4$ total holes.
the spatial tunneling distribution at the first electron excitation state, labeled as bc in Fig.~\ref{figS3} (a). Figure~\ref{figS3} (b-c) shows the intensity of tunneling DOS at
each spatial site without and with the inclusion of the tunneling matrix effect. In this case we instead find the spatial distribution has an internal $2\sqrt{2}a$
structure along the direction connecting the two Cu centers.
\par\endgroup
\end{minipage}

\clearpage
\noindent
\begin{minipage}[t]{0.48\textwidth}
\vspace{0pt}
\makeatletter\def\@captype{figure}\makeatother
\centering
\includegraphics[width=3.1in]{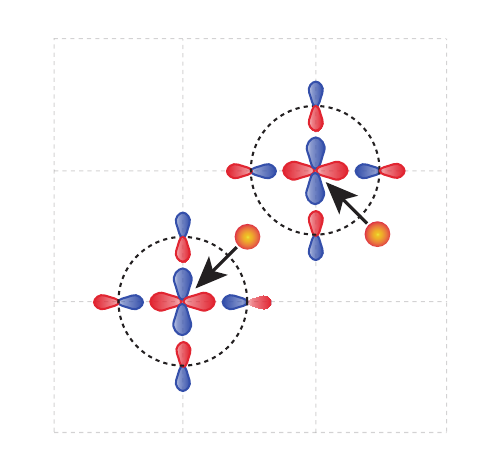}
		\caption{
The schematic of the 10-site Cu-O plaquette.
			\label{figS2}}
	
\end{minipage}\hfill
\begin{minipage}[t]{0.48\textwidth}
\vspace{0pt}
\makeatletter\def\@captype{figure}\makeatother
\centering
\includegraphics[width=3.1in]{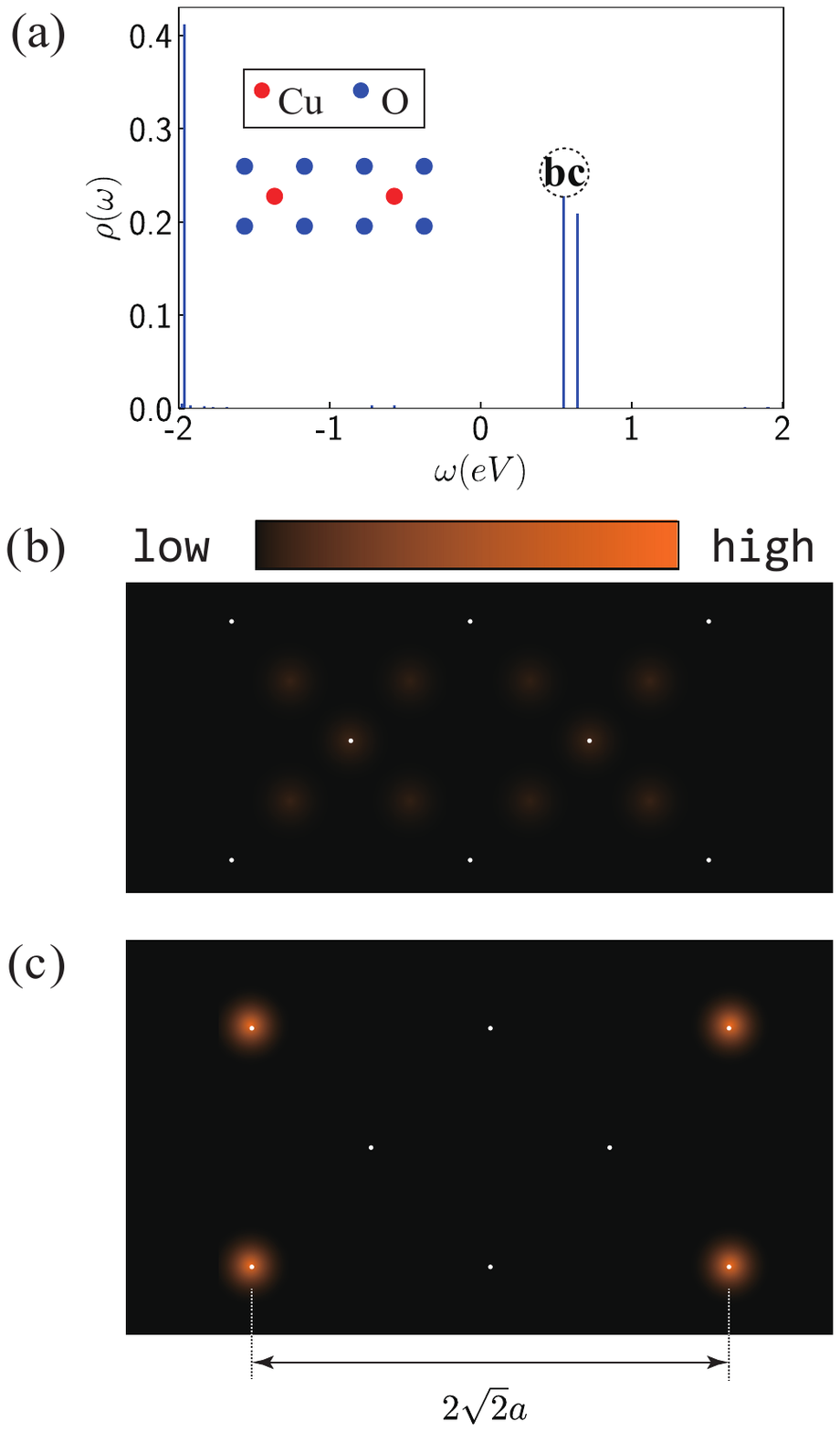}
		\caption{
(a) The tunneling density of states (DOS), $\rho(\omega)$, calculated for a two-hole doping scenario in the two-adjacent-Na-dopant situation, is plotted as a function of energy, $\omega$. The inset provides a schematic of the 10-site Cu-O plaquette, with red circles indicating Cu atoms and blue circles representing O atoms.
(b-c) Local density of states for the in-gap bound state identified in (a) are shown both without (b) and with (c) the inclusion of the tunneling matrix effect. The white dot pattern signifies the positions of Cu sites in CCOC, analogous to Cl atom locations. In (b), the DOS map assumes direct coupling of the STM probe to Cu or O atoms. Notably, in (c), the STM map evaluated above the Cu sites displays a stripe pattern with an approximate width of 2$\sqrt{2}a_0$.
			\label{figS3}}
	
\end{minipage}

\end{document}